\documentclass[aps,prl,twocolumn,groupedaddress,amsmath,showpacs]{revtex4}
\usepackage{amsmath}
\usepackage{graphicx}
\usepackage{amssymb}
\usepackage{hyperref}
\usepackage{color}
\usepackage{times,txfonts}
\usepackage{isomath}

\newcommand{\ket}[1]{|#1\rangle}
\newcommand{\bra}[1]{\langle#1|}

\begin{document}
\title{Quantification of Gaussian Quantum Steering}

\author{Ioannis Kogias}
\email{john$_$k$_$423@yahoo.gr}
\affiliation{$\mbox{School of Mathematical Sciences, The University of Nottingham,
University Park, Nottingham NG7 2RD, United Kingdom}$}

\author{Antony R. Lee}
\email{pmaal1@nottingham.ac.uk}
\affiliation{$\mbox{School of Mathematical Sciences, The University of Nottingham,
University Park, Nottingham NG7 2RD, United Kingdom}$}

\author{Sammy Ragy}
\email{pmxsr3@nottingham.ac.uk}
\affiliation{$\mbox{School of Mathematical Sciences, The University of Nottingham,
University Park, Nottingham NG7 2RD, United Kingdom}$}

\author{Gerardo Adesso}
\email{Gerardo.Adesso@nottingham.ac.uk}
\affiliation{$\mbox{School of Mathematical Sciences, The University of Nottingham,
University Park, Nottingham NG7 2RD, United Kingdom}$}

\begin{abstract}
Einstein-Podolsky-Rosen steering incarnates a useful nonclassical correlation which sits between entanglement and Bell nonlocality. While a number of qualitative steering criteria exist, very little has been achieved for what concerns quantifying steerability. We introduce a computable measure of steering for arbitrary bipartite Gaussian states of continuous variable systems. For two-mode Gaussian states, the measure reduces to a form of coherent information, which is proven never to exceed entanglement, and to reduce to it on pure states. We provide an operational connection between our measure and the key rate in one-sided device-independent quantum key distribution.
We further prove that Peres' conjecture holds in its stronger form within the fully Gaussian regime: namely, steering bound entangled Gaussian states by Gaussian measurements is impossible.
\end{abstract}


\date{December  18, 2014}
\maketitle

Quantum correlations have been intensively investigated in recent years after the realization that, besides their foundational importance, they can be exploited to outperform any classical approach in certain tasks, e.g.~in computation \cite{chuaniels}, secure communication \cite{ekert,crypto} and metrology \cite{metro}.  For mixed states of composite quantum systems, quantum correlations can manifest in different forms \cite{disc}. While entanglement \cite{ent} and Bell nonlocality \cite{nonloc} are two of the most well-studied such manifestations, an intermediate type of quantum correlation, known as quantum {\it steering} \cite{schr,schr2}, has only quite recently attracted a renewed interest from the quantum information community \cite{wiseman,Skrzypczyk}, opening new avenues for theoretical exploration and practical applications.

Steering is the quantum mechanical phenomenon that allows one party, Alice, to change (i.e.~to `steer') the state of a distant party, Bob, by exploiting their shared entanglement. This phenomenon, fascinatingly discussed by Schr\"odinger \cite{schr,schr2}, was already noted by Einstein, Podolksy and Rosen (EPR) in their famous 1935 paper \cite{epr}, and is at the heart of the so-called  EPR paradox \cite{eprpar}. There it was argued that steering implied an unacceptable ``action at a distance'', which led EPR to claim the incompleteness of quantum theory. The EPR expectations for local realism were mostly extinguished by Bell's theorems \cite{bell64, bell76}, which showed that no locally causal theory can reproduce all the correlations observed in nature \cite{aspect}. The first experimental criterion for the demonstration of the EPR paradox, i.e., for the detection of quantum steering, was later proposed by Reid \cite{reid}, but it was not until 2007 that the particular type of nonlocality captured by the concept of steering  \cite{epr,schr,schr2} was in fact formalized \cite{wiseman,wisepra}.

From a quantum information perspective \cite{wiseman}, steering corresponds to the task of verifiable entanglement distribution by an untrusted party. If Alice and Bob share a state which is steerable in one way, say from Alice to Bob, then Alice is able to convince Bob (who does not trust Alice) that their shared state is entangled, by performing local measurements and classical communication \cite{wiseman}. Notice that steering, unlike entanglement, is an asymmetric property: a quantum state may be steerable from Alice to Bob, but not vice versa. On the operational side, it has been recently realized that steering provides security in one-sided device-independent quantum key distribution (QKD) \cite{branciard}, where the measurement apparatus of one party only is untrusted. These protocols are less demanding than totally device-independent ones, for which Bell nonlocality is known to be necessary \cite{crypto}. Experimentally, at variance with the case of Bell tests, a demonstration of steering free of detection and locality loopholes is in reach \cite{branciard, vallone,wittmann,prydenew}, which makes one-sided device-independent QKD appealing for current technology and quantum steering a practically useful concept.  EPR steering also plays an operative role in channel discrimination \cite{PianiSt} and tele-amplification \cite{henewnew}.

Several experiments have been already performed, demonstrating steering and its asymmetry \cite{saunders,eberle,handchen,bennet,wittmann,smith,steinlechner,prydenew,chinagame}, and a number of recent studies have been devoted to improve our understanding of quantum steerability, ranging from the development of better criteria to detect steerable states \cite{cavalcanti,walborn11,walborn13,cavalcanti2,pramanik}, to the analysis of the distribution of steering among multiple parties \cite{asymmetry,he,reid13,bowles}. However unlike entanglement, for which a variety of operationally-motivated measures exist \cite{ent,virplenio}, there is still a surprisingly scarce literature addressing the fundamental question of {\it quantifying} how steerable a given quantum state is \cite{Skrzypczyk,PianiSt,angelo}.

In this Letter we present a comprehensive quantitative investigation of steerability in the archetypical setting of bipartite continuous variable systems, for which the very notion of EPR steering was originally debated and analyzed \cite{epr,reid}. We focus on a fully Gaussian scenario: namely, we consider generally mixed multimode bipartite Gaussian states, that constitute a distinctive corner of the infinite-dimensional Hilbert space \cite{ourreview,ournewreview,weedbrook}, and study their steerability under Gaussian measurements \cite{gaussian1,gaussian2}. By analyzing the degree of violation of a necessary and sufficient criterion for Gaussian steerability \cite{wiseman,wisepra}, we obtain a {\it computable} measure of Gaussian steering, and we investigate its properties. In the special case of two-mode Gaussian states, we characterize the maximum allowed steering asymmetry, we connect the measure operationally to the key rate of one-sided device-independent QKD \cite{walk},
 and we show that the Gaussian steering degree is upper bounded by the Gaussian R\'enyi-$2$ entanglement \cite{renyi}, with equality on pure states.
 Finally, we prove in general that (multimode) bound entangled Gaussian states cannot be steered by Gaussian measurements, a result of relevance in view of the recent debate about a conjecture by Peres and its recently proposed strenghtening by Pusey \cite{peres,pusey,puseywrong,pereswrong}.


Let us first briefly introduce the reader to the Gaussian realm. Gaussian states and operations play a central role in the analysis and implementation of continuous variable quantum technologies \cite{ourreview,weedbrook}. In our Letter, we consider a generic Gaussian  $(n+m)$-mode state ${\rho _{AB}}$ of a bipartite system, comprised of a subsystem $A$ (for Alice) of $n$ modes and a subsystem $B$ (for Bob) of $m$ modes. For each mode $j$, belonging to $A$ ($B$), we define the phase-space operators $\hat x_j^{A(B)},\,\,\hat p_j^{A(B)}$, grouped for convenience into the vector $\hat R = (\hat x_1^A,\hat p_1^A, \ldots ,\hat x_n^A,\hat p_n^A,\hat x_1^B,\hat p_1^B, \ldots ,\hat x_m^B,\hat p_m^B)^{\sf T}$, satisfying the canonical commutation relations which are compactly written as $[{{{\hat R}_i},{{\hat R}_j}} ] = i{\Omega _{ij}}$, with $\Omega  =  \bigoplus _1^{n+m} {{\ 0\ \ 1}\choose{-1\ 0}}$ being the symplectic form. Any Gaussian state ${\rho _{AB}}$ is fully specified, up to local displacements, by its covariance matrix (CM),
\begin{equation}\label{CM}
\sigma_{AB} = \left( {\begin{array}{*{20}{c}}
   A & C  \\
   {{C^{\sf T}}} & B  \\
\end{array}} \right),
\end{equation} with elements ${\sigma _{ij}} = \text{Tr}\big[ {{{\{ {{{\hat R}_i},{{\hat R}_j}} \}}_ + }\ {\rho _{AB}}} \big]$. Notice that the submatrices $A$ and $B$ are the CMs correspoding to the reduced states of Alice's and Bob's subsystems respectively. Every CM $\sigma_{AB}$ that corresponds to a physical quantum state has to satisfy the \textit{bona fide} condition,
\begin{equation}\label{bonafide}
{\sigma _{AB}} + i\,({\Omega _A} \oplus {\Omega _B}) \ge 0.
\end{equation}

Let us now formally define steerability. Under a set of measurements $\mathcal{M}_A$ on Alice, a bipartite state $\rho_{AB}$ is $A\to B$ steerable---i.e., Alice can steer Bob---\textit{iff} it is \textit{not} possible
for every pair of local observables $R_A \in \mathcal{M}_A$ on $A$ and $R_B$ (arbitrary) on $B$, with respective outcomes $r_A$ and $r_B$, to express the joint probability as   \cite{wiseman}
$P\left( {{r_A},{r_B}|{R_A},{R_B},{\rho _{AB}}} \right) = \sum\limits_\lambda  {{\wp_\lambda }} \, \wp\left( {{r_A}|{R_A},\lambda } \right)P\left( {{r_B}|{R_B},{\rho _\lambda }} \right)$. That is, at least one measurement pair $R_A$ and $R_B$ must violate this expression when ${\wp_\lambda }$ is fixed across all measurements. Here ${\wp_\lambda }$ and $\wp \left( {{r_A}|{R_A},\lambda }\right)$ are arbitrary probability distributions and $P\left( {{r_B}|{R_B},{\rho _\lambda }} \right)$ is a probability distribution subject to the extra condition of being evaluated on a quantum state $\rho_\lambda$, meaning that a complete knowledge of Bob's devices (but not of Alice's ones) is required to formulate the steering condition.

In the fully Gaussian scenario, where  $\rho_{AB}$ is a Gaussian state described by the CM $\sigma_{AB}$ \eqref{CM}, we will focus on Alice's measurement set $\mathcal{M}_A$ to be Gaussian (i.e., mapping Gaussian states into Gaussian states). A Gaussian measurement \cite{gaussian2}, which is generally implemented via symplectic transformations followed by balanced homodyne detection, can be described by a positive operator with a CM $T^{{R_A}}$, satisfying ${T^{{R_A}}} + i\,{\Omega _A} \ge 0$. Every time Alice makes a measurement $R_A$ and gets an outcome $r_A$, Bob's conditioned state $\rho_{B}^{r_A|R_A}$ is Gaussian with a CM given by  $B_{}^{{R_A}} = B - C{\left( {{T^{{R_A}}} + A} \right)^{ - 1}}{C^{\sf T}}$, independent of Alice's outcome.

It can be shown \cite{wiseman} that a general $(n+m)$-mode Gaussian state $\rho_{AB}$ is $A\to B$ steerable by Alice's Gaussian measurements \textit{iff} the condition
\begin{equation}\label{nonsteer}
{\sigma _{AB}} + i\,({0_A} \oplus {\Omega _B}) \ge 0,
\end{equation}
is violated. 
Writing this in matrix form, using (\ref{CM}), the nonsteerability inequality (\ref{nonsteer}) is equivalent to two simultaneous conditions: (i) $A > 0$, and (ii) ${M^B_{\sigma}} + i{\Omega _B} \ge 0$, where $M^B_{\sigma} = B - {C^{\sf T}}{A^{ - 1}}C$ is the Schur complement of $A$ in the CM $\sigma_{AB}$.
Condition (i) is always verified since $A$ is a physical CM. Therefore, $\sigma_{AB}$ is $A \to B$ steerable \textit{iff} the symmetric and positive definite $2m \times 2m$  matrix $M^B_{\sigma}$ is not a {\it bona fide} CM, i.e., if condition (ii) is violated \cite{wiseman,wisepra}.
By Williamson's theorem \cite{williamson},  $M^B_{\sigma}$ can be diagonalized by a symplectic transformation $S_B$ such that $S_B M^B_{\sigma} S_B^{\sf T}=\text{diag}\{\bar{\nu}^B_1,\bar {\nu}^B_1,\ldots,\bar{\nu}^B_m,\bar {\nu}^B_m\}$, where $\{\bar{\nu}^B_{j}\}$ are the symplectic eigenvalues of $M^B_{\sigma}$, which can be determined by $m$ local symplectic invariants \cite{seraleprl}; alternatively, they can be computed as the orthogonal eigenvalues of the matrix $|i \Omega_B M^B_{\sigma}|$. The nonsteerability condition (\ref{nonsteer}) is thus equivalent to $\bar{\nu}^B_j \geq 1$ for all $j=1,\ldots,m$.

We then propose to {\it quantify} how much a a bipartite $(m+n)$-mode Gaussian state with CM $\sigma_{AB}$ is steerable (by Gaussian measurements on Alice's side) via the  following quantity
\begin{equation}\label{GSAB}
{\cal G}^{A \to B}(\sigma_{AB}):=
\max\bigg\{0,\,-\sum_{j:\bar{\nu}^B_j<1} \ln(\bar{\nu}^B_j)\bigg\}\,.
\end{equation}
This quantity, hereby defined as  Gaussian $A \to B$ steerability, is invariant under local unitaries (symplectic operations at the CM level), it vanishes \textit{iff} the state described by $\sigma_{AB}$ is nonsteerable by Gaussian measurements, and it generally quantifies the amount by which the condition (\ref{nonsteer}) fails to be fulfilled. Clearly, a corresponding measure of Gaussian $B \to A$ steerability can be obtained by swapping the roles of $A$ and $B$, resulting in an expression like (\ref{GSAB}), in which the symplectic eigenvalues of the $2n \times 2n$ Schur complement of $B$,   $M^A_{\sigma} =  A - {C}{B^{ - 1}}{C^{\sf T}}$,  appear instead.  We highlight the formal similarity with the formula for the logarithmic negativity \cite{vidalwerner,plenioprl,ent,virplenio}---an entanglement measure which quantifies how much the positivity of the partial transpose condition for separability is violated \cite{peres96,horodecki96,simon00,werewolf}---for Gaussian states; in the latter case, however, the symplectic eigenvalues of the partially transposed CM are considered \cite{virplenio,vidalwerner,extremal,ourreview}.

The proposed measure of steering is easily computable for bipartite Gaussian states of an arbitrary number of modes. When the steered party, e.g.~Bob in Eq.~(\ref{GSAB}), has one mode only ($m=1$), the Gaussian steerability acquires a particularly simple form. Indeed, in such a case, $M^B_{\sigma}$ has a single symplectic eigenvalue, $\bar{\nu}^B = \sqrt{\det M^B_{\sigma}}$; recalling that, by definition of Schur complement, $\det{\sigma_{AB}} = \det{A} \det{M^B_{\sigma}}$, we have
\begin{equation}
{\cal G}^{A \to B}(\sigma_{AB}) =
\mbox{$\max\big\{0,\, \frac12 \ln {\frac{\det A}{\det \sigma_{AB}}}\big\}$}= \max\big\{0,\, {\cal S}(A) - {\cal S}(\sigma_{AB})\big\}\,, \label{GS1}
\end{equation}
where we have introduced the R\'enyi-$2$ entropy ${\cal S}$, which for a Gaussian state with CM $\sigma$ reads ${\cal S}(\sigma) = \frac12 \ln( \det \sigma)$ \cite{renyi}.

Interestingly, the quantity ${\cal S}(A) - {\cal S}(\sigma_{AB}) \equiv {\cal I}^{A \langle B}$ can be seen as a form of quantum {\it coherent information} \cite{wilde}, but with R\'enyi-$2$ entropies replacing the conventional von Neumann entropies.
Thanks to this connection, we can now prove some valuable properties of the Gaussian steering measure (\ref{GS1}) for $(n+1)$-mode Gaussian states, namely: (a) ${\cal G}^{A \to B}$ is convex; (b) ${\cal G}^{A \to B}$ is monotonically decreasing under quantum operations on the (untrusted) steering party Alice; (c) ${\cal G}^{A \to B}$ is additive, i.e., given the tensor product of two Gaussian states $\rho_{AB}^1 \otimes \rho_{AB}^2$ with corresponding CM $\sigma_{AB}^1 \oplus \sigma_{AB}^2$, then ${\cal G}^{A \to B}(\sigma^1_{AB} \oplus \sigma^2_{AB}) = {\cal G}^{A \to B}(\sigma^1_{AB})+ {\cal G}^{A \to B} (\sigma^2_{AB})$;  (d) ${\cal G}^{A \to B}(\sigma_{AB}) = {\cal E}(\sigma^p_{AB})$ for $\sigma^p_{AB}$ pure, and (e) ${\cal G}^{A \to B}(\sigma_{AB}) \leq {\cal E}(\sigma_{AB})$ for $\sigma_{AB}$ mixed, where  ${\cal E}$ denotes the Gaussian R\'enyi-$2$  measure of entanglement \cite{renyi}.
The proof of (a) follows from the concavity of the R\'enyi-$2$ entropy. The proof of (b) follows from the fact that the Gaussian R\'enyi-$2$ coherent information ${\cal I}^{A \langle B}$ obeys the data processing inequality  (which in turn is a consequence of the strong subadditivity of the R\'enyi-$2$  entropy ${\cal S}$ for Gaussian states) \cite{renyi,wilde}, ${\cal I}^{A' \langle B} \leq {\cal I}^{A \langle B}$ if $A'$ is obtained from $A$ by the action of a Gaussian quantum channel. Property (c) follows from straightforward linear algebra and the additivity of the logarithm. The proof of (d) is immediate, as for pure states ${\cal S}(\sigma^p_{AB})=0$ and ${\cal E}(\sigma^p_{AB}) = {\cal S}(A)$. Property (e) needs to be proven when  ${\cal G}^{A \to B}>0$, in which case ${\cal G}^{A \to B}={\cal I}^{A \langle B}$. We recall that the R\'enyi-$2$ entanglement of a bipartite Gaussian state $\rho_{AB}$ is defined via a Gaussian convex roof procedure \cite{ourreview,renyi}, ${\cal E}(\rho_{AB}) = \inf_{\{p_i,\,\ket{\psi_i}\}} \sum_i p_i {\cal S}(\text{Tr}_{B} \ket{\psi_i}\bra{\psi_i})$, where the pure states $\{\ket{\psi_i}\}$ are Gaussian; let us denote by $\{p'_i, \ket{\psi'_i}\}$ the optimal decomposition of $\rho_{AB}$ which minimizes the R\'enyi-$2$ entanglement. We have then ${\cal E}(\rho_{AB}) =  \sum_i p'_i {\cal S}(\text{Tr}_{B} \ket{\psi'_i}\bra{\psi'_i}) = \sum_i p'_i {\cal I}^{A \langle B}(\ket{\psi'_i}\bra{\psi'_i}) \geq {\cal I}^{A \langle B} (\sum_i p'_i  \ket{\psi'_i}\bra{\psi'_i}) = {\cal I}^{A \langle B}(\rho_{AB}) = {\cal G}^{A \to B}(\rho_{AB})$, where we used, in order, properties (d) and (a). Remarkably,  properties (d) and (e) demonstrate that our measure of Gaussian steering respects the hierarchy of quantum correlations \cite{wiseman}. Property (b) is interesting, and we leave it for further investigation to establish whether any valid measure of steering \cite{Skrzypczyk,PianiSt} should obey such monotonicity \cite{steerestheory}.

\begin{figure}[t!]
\includegraphics[width=6cm]{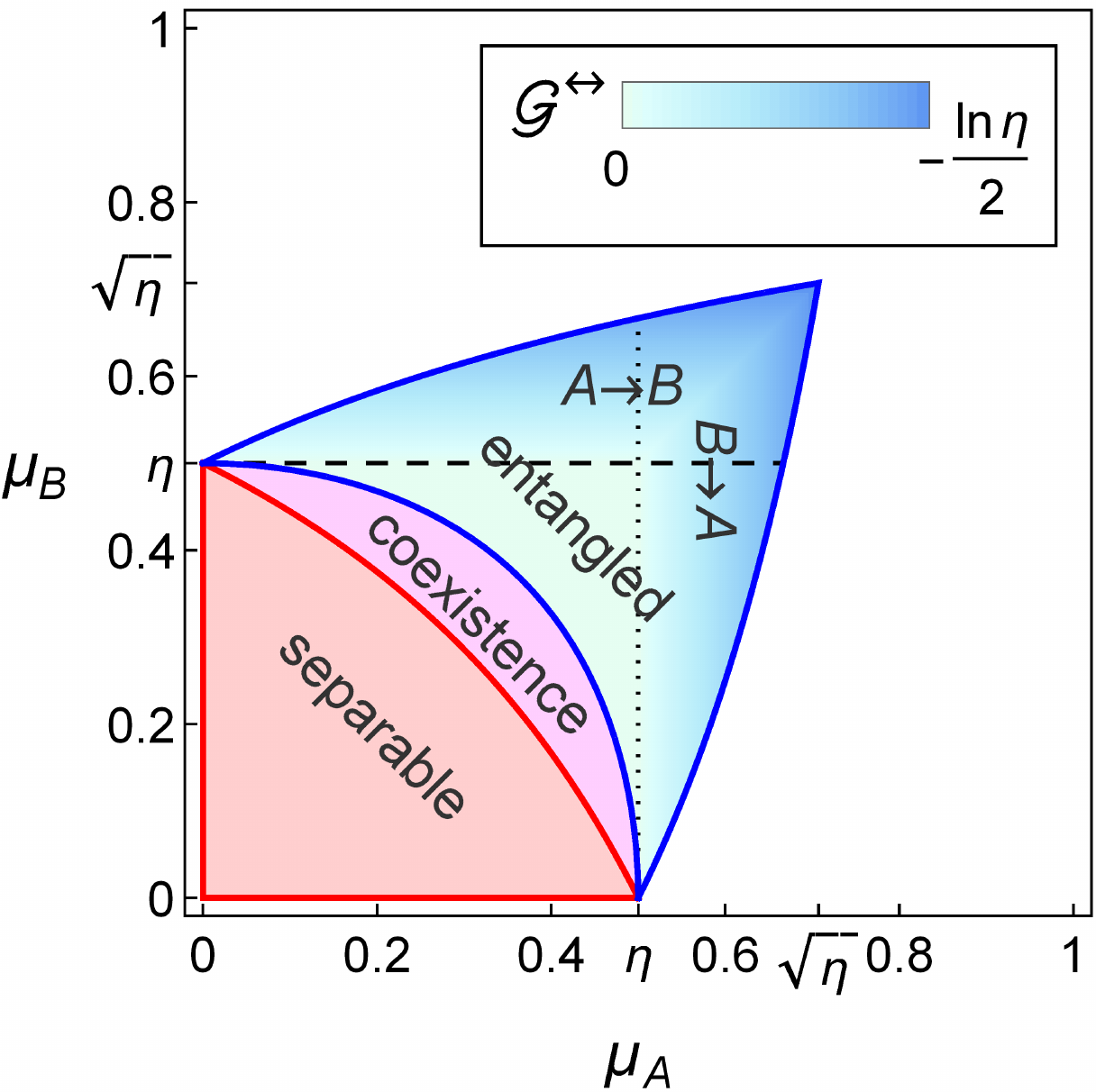}
\caption{(Color online) Classification of separability and Gaussian steerability of two-mode Gaussian states with marginal purities $\mu_A$ and $\mu_B$ and global purity $\mu = (\mu_A \mu_B)/\eta$, here plotted for $\eta=\frac12$. By Gaussian measurements, states above the dashed line are $A \to B$ steerable and states to the right of the dotted line are $B \rightarrow A$ steerable. An overlay of the symmetrized degree of steerability ${\cal G}^{\leftrightarrow} \equiv \max\{{\cal G}^{A \to B},\,{\cal G}^{B \to A}\}$ is depicted in the region of entangled states. See text for further details on the various regions and their boundaries.}
\label{ficaclassica}
\end{figure}

In the following, we specialize our attention onto the paradigmatic case of two-mode Gaussian states ($n=m=1$), for which the degree of steering in both ways can be easily measured according to our definition: ${\cal G}^{A \to B}(\sigma_{AB}) = \max\{0,\, {\cal S}(A) - {\cal S}(\sigma_{AB})\}$ and ${\cal G}^{B \to A}(\sigma_{AB}) = \max\{0,\, {\cal S}(B) - {\cal S}(\sigma_{AB})\}$. Qualification and quantification of steering in two-mode Gaussian states thus reduces entirely to an interplay between the global purity $\mu = 1/\sqrt{\det \sigma_{AB}}$ and the two marginal purities $\mu_{A(B)} = 1/\sqrt{\det A (B)}$. Introducing the ratio $\eta = (\mu_A \mu_B)/\mu$, all physical two-mode Gaussian states live in the region $\eta_0 \leq \eta \leq 1$ where $\eta_0=\mu_A \mu_B + |\mu_A - \mu_B|$ \cite{extremal}. States with $\eta_s \leq \eta \leq 1$ where $\eta_s=\mu_A + \mu_B - \mu_A \mu_B $ are necessarily separable, states with $\eta_e \leq \eta < \eta_s$ where $\eta_e=\sqrt{\mu_A^2+\mu_B^2-\mu_A^2\mu_B^2}$ can be entangled or separable (coexistence region), while states with $\eta_0 \leq \eta < \eta_e$ are necessarily entangled \cite{extremal}. Within the latter region, states with $\eta \geq \{\mu_A, \mu_B\}$ are nonsteerable; states with $\eta < \mu_B$ are $A \to B$ steerable; states with $\eta < \mu_A$ are $B \to A$ steerable. This allows us to  classify the separability and steerability (by Gaussian measurements) of all two-mode Gaussian states in the $(\mu_A, \mu_B, \eta)$ space, completing the programme advanced a decade ago in~\cite{prl,extremal}. A cross-section of this insightful classification for $\eta=\frac12$ is visualized in Fig.~\ref{ficaclassica}.

\begin{figure}[t!]
\includegraphics[width=8cm]{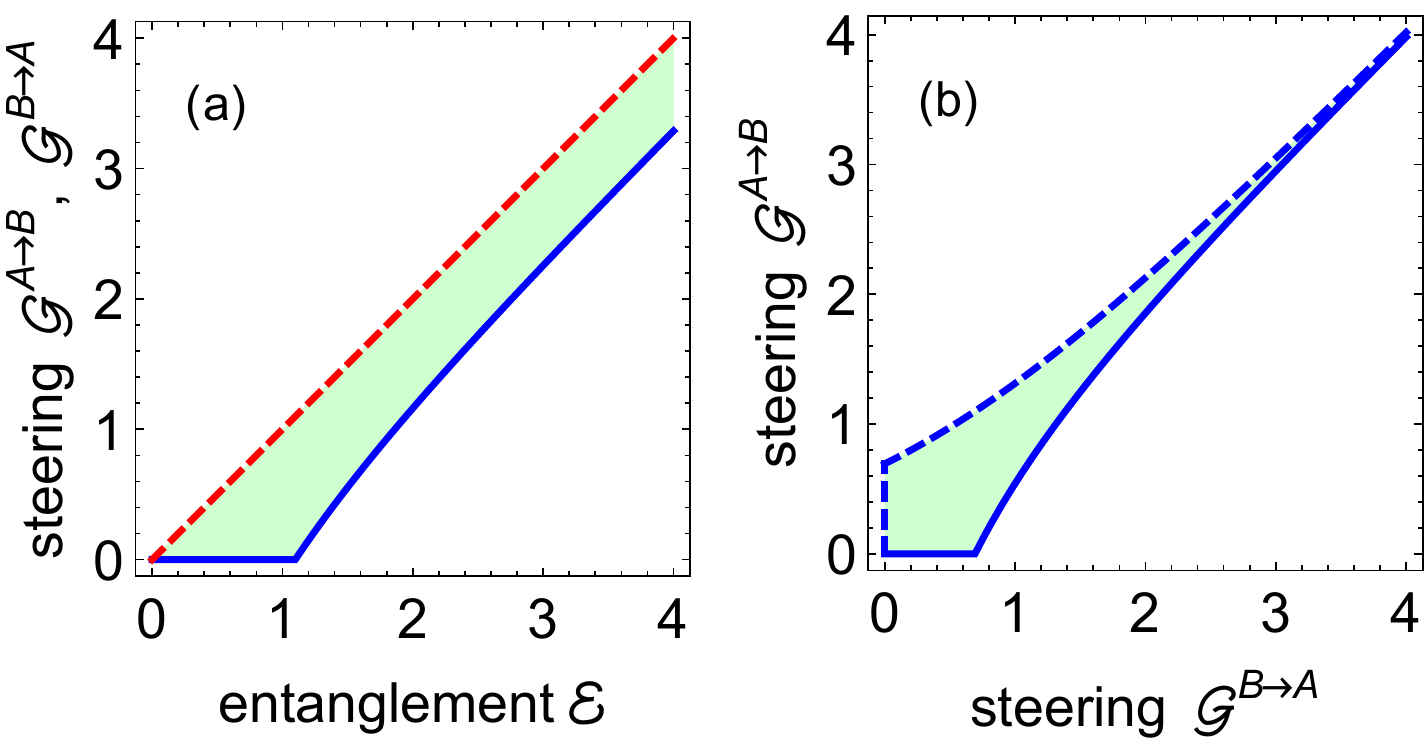}
\caption{(Color online) Plots of (a) Gaussian steerability versus Gaussian R\'enyi-$2$ entanglement and (b) $A \to B$ versus $B \to A$ Gaussian steerability, for two-mode Gaussian states. Physically allowed states fill the shaded (green online) regions. Pure states  $\sigma_{AB}^p$ sit on the upper (dashed) boundary in panel (a); the lower (solid) boundaries in both plots accommodate extremal states  $\sigma_{AB}^x$, while swapping $A$ and $B$ in them one obtains states $\sigma_{BA}^x$ which fill the upper boundary in (b).}
\label{ficasfondata}
\end{figure}

We have seen in general how steering can never exceed entanglement for Gaussian states (with one steered mode). It is interesting to investigate  how {\it small} ${\cal G}^{A \to B}$ can also be for a given R\'enyi-$2$ entanglement ${\cal E}$, on arbitrary two-mode Gaussian states. To address this question we exploit the local-unitary-invariance of  ${\cal G}^{A \to B}$, and consider without loss of generality its evaluation on CMs (\ref{CM}) in standard form, characterized by $A=\text{diag}(a,a)$, $B=\text{diag}(b,b)$, $C=\text{diag}(c,d)$. We can then perform a constrained minimization of ${\cal G}^{A \to B}$ at fixed ${\cal E}$, over the covariances $a,b,c,d$, subject to the condition (\ref{bonafide}). We find that the extremal states sit on the boundary $\eta = \eta_0$, and have a CM $\sigma_{AB}^x$ specified by $b=a-1+a/s$, $-d=c=\sqrt{(a-1)(s+1)(a/s)}$, with  $a \geq s \geq 1$, in the limit $a \rightarrow \infty$. For these extremal states, ${\cal G}^{A \to B}(\sigma_{AB}^x)=\ln (s)$ and ${\cal E}(\sigma_{AB}^x) = \ln (2s+1)$. Analogous results hold for  ${\cal G}^{B \to A}$. For all two-mode Gaussian states with a given ${\cal E}$, the steering measures thus admit an upper {\it and} a lower bound [see Fig.~\ref{ficasfondata}(a)],
$\max\left\{0,\,\ln \mbox{$\big[\frac12(e^{{\cal E}}-1)\big]$}\right\} \leq \big\{{\cal G}^{A \to B},\,{\cal G}^{B \to A}\big\} \leq {\cal E}$,
where the leftmost inequality is saturated on the extremal states $\sigma_{AB}^x$, and the rightmost one on pure (two-mode squeezed) states $\sigma_{AB}^p$, specified by $b=a, -d=c=\sqrt{a^2-1}$. This entails, in particular, that all two-mode Gaussian states with ${\cal E}> \ln 3 \approx 1.1$ are necessarily steerable in both ways; for highly entangled states, ${\cal E} \gg 0$, the Gaussian steering measure (in either way) remains bounded between ${\cal E}$ and ${\cal E} - \ln 2$.

The asymmetry of steering in the Gaussian setting has been experimentally demonstrated in \cite{handchen}. Clearly,   ${\cal G}^{A \to B} \neq {\cal G}^{B \to A}$ in general, but how asymmetric can steerability be, at most, on two-mode Gaussian states? By maximizing the difference  $|{\cal G}^{B \to A}-{\cal G}^{A \to B}|$ on standard form CMs, we find quite intriguingly that the states endowed with maximum steering asymmetry are exactly the ones with  CM $\sigma_{AB}^x$ defined above, for which ${\cal G}^{A \rightarrow B} = \ln (s)$ and ${\cal G}^{B \rightarrow A} = \ln (s+1)$. For all two-mode Gaussian states, one has then $\max\{0,\, \ln[\exp({\cal G}^{A \to B})-1]\} \leq {\cal G}^{B \to A} \leq \ln[\exp({\cal G}^{A \to B})+1]$. This entails that the steering asymmetry $|{\cal G}^{B \to A}-{\cal G}^{A \to B}|$ can never exceed $\ln 2$, it is maximal when the state is nonsteerable in one way, and it decreases with increasing steerability in either way [see Fig.~\ref{ficasfondata}(b)].

We now investigate operational interpretations for the proposed steering quantifier(s) for two-mode Gaussian states. We observe from \cite{wiseman,wisepra} that our measures, evaluated on standard form CMs, are monotonic functions of the product of the (minimum) conditional variances associated to local homodyne detections, which appear in the seminal Reid criterion for the EPR paradox \cite{reid}: namely,  ${V_{{x_A}|{x_B}}}{V_{{p_A}|{p_B}}} = \det {M^A_{\sigma}} = \det{\sigma_{AB}}/\det B$
 and ${V_{{x_B}|{x_A}}}{V_{{p_B}|{p_A}}} = \det {M^B_{\sigma}} = \det{\sigma_{AB}}/\det A$; this renders   ${\cal G}^{A \to B}$ and ${\cal G}^{B \to A}$  directly accessible experimentally. We can then show that these measures find important applications for the task of one-sided device-independent QKD \cite{branciard}, which has been recently extended to continuous variables \cite{walk}. Considering the relevant entanglement-based protocol \cite{ekert}, let a two-mode entangled Gaussian state with CM $\sigma_{AB}$ in standard form be shared between Alice and Bob, who want to establish a secret key. By performing homodyne detections on their modes, and a direct reconciliation scheme (where Alice sends corrections to Bob), they can achieve a secret key rate
$K \ge \max \bigg\{ {0,\,\ln \bigg( {\frac{2}{{e\sqrt {{V_{{x_A}|{x_B}}}{V_{{p_A}|{p_B}}}} }}} \bigg)} \bigg\}$ \cite{walk}. This bound can be readily expressed  in terms of the $B \to A$ Gaussian steerability of $\sigma_{AB}$, yielding
\begin{equation}\label{key}
K \ge \max \{0,\, {\cal G}^{B \to A}(\sigma_{AB}) + \ln 2 -1\}.
\end{equation}
In the case of a reverse reconciliation protocol, the corresponding key rate \eqref{key} would involve ${\cal G}^{A \rightarrow B}$ rather than ${\cal G}^{B \rightarrow A}$.  Therefore, the degree of Gaussian steerability defined in this Letter nicely quantifies the  guaranteed key rate achievable within a practical semi-device-independent QKD setting, realizable with current optical technology \cite{handchen,walk}.


Finally, we address the more fundamental question of steerability of bound entangled Gaussian states. Peres conjectured that states whose entanglement cannot be distilled, i.e., bound entangled states \cite{ent}, cannot violate any Bell inequality \cite{peres}. Recently, Pusey proposed a stronger conjecture, namely that bound entangled states cannot even display  EPR steering \cite{pusey}. Surprisingly, both  conjectures have been now disproven, by identifying steerable \cite{puseywrong} and nonlocal \cite{pereswrong} bound entangled qudit states. However, the question stayed open for continuous variable systems, and we settle it  in the Gaussian case. Let  $\sigma_{AB}$ be the CM of a general bound entangled $(n+m)$-mode Gaussian state. Any such state obeys the {\it bona fide} condition (\ref{bonafide}) as well as the condition
\begin{equation}
{\sigma _{AB}} + i\left( {\, - {\Omega _A}} \right) \oplus {\Omega _B} \ge 0\,,
 \end{equation}
 which amounts to positivity under partial transposition \cite{werewolf}. Adding the two matrix inequalities together, one obtains (twice) the nonsteerability condition (\ref{nonsteer}). This remarkably simple proof yields a general no-go result: steering bound entangled Gaussian states by Gaussian measurements is impossible; i.e., the Peres-Pusey conjecture holds in a fully Gaussian scenario.

In conclusion, we presented an intuitive and computable quantification of EPR steering \cite{wiseman} for bipartite Gaussian states under Gaussian measurements. We linked our measure to the key rate of one-sided device-independent QKD \cite{walk} and proved hierarchical relationships with entanglement.
This Letter delivers substantial advances for the characterization of EPR steering and provides an important addition to the  established framework of Gaussian quantum information theory \cite{weedbrook,ourreview,ournewreview}. In principle, our approach might be applied as well to general states: Namely, for a (non-Gaussian) bipartite state $\rho_{AB}$, one can define an indicator of steerability by Gaussian measurements as in Eq.~(\ref{GSAB}), with $\sigma_{AB}$ denoting the CM of the second moments of $\rho_{AB}$. This may be connected, in general, to the degree of violation of optimized Reid--type linear variance criteria for EPR steering
\cite{reid,wisepra,cavalcanti,eprpar,steerepaps}.
Notice however that a bipartite non-Gaussian state $\rho_{AB}$ can still be steerable even if its ${\cal G}^{A \to B}$ vanishes, as the state may possess EPR correlations only detectable via nonlinear criteria involving higher order moments \cite{cavalcanti,walborn11}; e.g., a two-qubit pure Bell state is clearly steerable but its CM fails to violate (\ref{nonsteer}).

The interplay between EPR steering \cite{wiseman}, `obesity' of steering ellipsoids \cite{ellipsoid}, and other forms of asymmetric nonclassical correlations such as discord  \cite{disc,adessodatta,giordaparis,henew}, is worthy of further investigation. We also plan to generalize our  analysis to multipartite settings \cite{he}, in order to derive quantitative  monogamy inequalities for steering \cite{reid13}, complementing the existing ones for Gaussian entanglement \cite{ourreview,strongmono,renyi}.

We acknowledge fruitful discussions with A. Doherty, Q. Y. He, M. Piani, T. C. Ralph, M. D. Reid, D. Cavalcanti, E. G. Cavalcanti, P. Skrzypczyk, F. Toscano, S. Walborn, and H. Wiseman. We thank the University of Nottingham (International Collaboration Fund), the EPSRC (Doctoral Prize), the Foundational Questions Institute (FQXi-RFP3-1317) and the Brazilian CAPES (Grant~No.~108/2012) for financial support.

\medskip

\noindent {\it Note added.} After completion of the present manuscript, an independent proof of Peres' conjecture for Gaussian states under Gaussian measurements has been reported in Ref.~\cite{pigliu}.


\end{document}